\documentclass[prd,tightenlines,11pt,onecolumn,amsmath,amssymb,showpacs]{revtex4}
\usepackage{graphicx,scalefnt,rotating}
\newcommand{\abbrev}{\scalefont{.9}}

\newcommand{\eqn}[1]{Eq.\,(\ref{#1})}
\newcommand{\fig}[1]{Fig.\,\ref{#1}}

\newcommand{\sct}[1]{Sect.\,\ref{#1}}

\newcommand{\order}[1]{{\cal O}(#1)}
\newcommand{\lo}{{\abbrev LO}}
\newcommand{\nlo}{{\abbrev NLO}}
\newcommand{\nnlo}{{\abbrev NNLO}}

\newcommand{\wbf}{{\abbrev WBF}}
\newcommand{\lhc}{{\abbrev LHC}}
\newcommand{\qcd}{{\abbrev QCD}}
\newcommand{\dis}{{\abbrev DIS}}

\newcommand{\sql}{{\abbrev SQL}}

\newcommand{\GeV}{\unskip\,\mathrm{GeV}}

\def\readRCS$#1: #2,v #3 #4 #5${%
 \def\filename{#2}%
 \def\fileversion{#3}%
 \def\filedate{#4}%
}

\readRCS
 $Id: ggwbf.tex,v 1.5 2007/12/14 11:35:36 rharland Exp $

\begin{document}

\title{\vspace*{-6em}
    \begin{flushright}
      \sf January 2008 --- NIKHEF/2007-025, WUB/07-12
    \end{flushright}
    \vspace*{3em}
    Gluon-Induced Weak Boson Fusion}
  \author{Robert V. Harlander$^{(1)}$, 
Jens Vollinga$^{(2)}$ and
Marcus M. Weber$^{(3)}$\\[1em]
  {\it
    $(1)$ Fachbereich C, 
    Bergische Universit\"at Wuppertal, D-42097 Wuppertal,
    Germany\\[-.3em]
    {\footnotesize\tt robert.harlander@uni-wuppertal.de}\\
    $(2)$ Theoretical Physics, NIKHEF, 1098 SJ Amsterdam, The
    Netherlands\\[-.3em]
    {\footnotesize\tt jensv@nikhef.nl}\\
    $(3)$ Department of Physics, University at Buffalo,
    Buffalo, NY 14260-1500,  USA\\[-.3em]
    {\footnotesize\tt mmweber@buffalo.edu}
  }}

\begin{abstract}
The gluon-gluon induced terms for Higgs production through weak boson
fusion (\wbf{}) are computed. Formally, these are of \nnlo{} in the
strong coupling constant. This is the lowest order at which non-zero
color exchange occurs between the scattering quarks, leading to a color
field and thus additional hadronic activity between the outgoing
jets. Using a minimal set of cuts, the numerical impact of these terms
is at the percent level with respect to the \nlo{} rate for weak boson
fusion. Applying the so-called \wbf{} cuts leads to an even stronger
suppression, so that we do not expect a significant deterioration of the
\wbf{} signal by these color exchange effects.
\end{abstract}


\maketitle

\section{Introduction}

One of the primary goals of the {\abbrev CERN} Large Hadron Collider
(\lhc{}) is to find the mechanism for electro-weak symmetry breaking.
The ``standard'' solution suggested many years ago predicts a
fundamental scalar particle, called the Higgs boson (for reviews, see
Refs.\,\cite{hunter,Djouadi:2005gi,Djouadi:2005gj}).  The interplay of
extremely precise measurements and similarly precise calculations allows
one to restrict its mass to the range between 114 and about
200\,GeV~\cite{lepewwg} which is quite remarkable considering the fact
that, a priori, this quantity is a free parameter of the theory.
Unfortunately, the mass region below around 140\,GeV turns out to be
rather problematic for Higgs discovery at a hadron collider. The reason
is that Higgs decay into weak gauge bosons is kinematically strongly
suppressed, and the Higgs branching ratio into $b\bar b$ becomes
dominant, reaching 70\% at $M_H=120$\,GeV and more than 80\% below
$M_H=100$\,GeV.  Without any additional tags from the production
process, the signal is then completely swamped by the \qcd{} production
of $b\bar b$ pairs.  Gluon fusion, which is the Higgs production process
with the largest cross section at the \lhc{}, does not provide such
additional tags which is why one has to use it in combination with the
rare decay into photons.

A few years ago it was pointed out that weak boson fusion (\wbf{}) is
very well suited for Higgs discoveries in the low mass region, provided
that the kinematical distribution of the outgoing jets is properly
identified~\cite{Rainwater:1997dg}. The distinguishing feature of this
process is the $t$-channel exchange of only color-less objects (weak
gauge bosons), from which the Higgs boson is radiated. The signal
therefore consists of two hard jets which can be found in opposite
hemispheres of the detector at large rapidities, while the Higgs decay
products are found at more central rapidities, without much more
additional hadronic activity.

Even at next-to-leading order (\nlo{}), color exchange among the
scattering partons is strongly suppressed due to the fact that it
requires a $t$-$u$-channel
interference~\cite{Han:1992hr,Spira:1997dg,Djouadi:1999ht,Figy:2003nv,
Berger:2004pc,Ciccolini:2007jr:2007ec}. 
Thus, the next-to-next-to-leading order (\nnlo{}) is the lowest order at
which true color-exchange diagrams can contribute significantly to the
\wbf{} signal.  The interesting issue is whether the jets produced by
such a mechanism have similar kinematical distributions as the typical
\wbf{} jets.  In this case, the color exchange might lead to additional
soft radiation which could deteriorate the \wbf{} signal. 

Note that there is a process with the same final state $H+2$jets that
does not involve weak gauge bosons: it is due to sub-processes like
$gg\to Hgg$, for example, where the gluons couple to the Higgs boson via
a top quark loop. Such terms are part of the \nnlo{} contribution to
gluon fusion~\cite{Harlander:2002wh,Anastasiou:2002yz,Ravindran:2003um}
and must be considered as background when aiming for \wbf{}. They were
calculated at leading order (\lo{}) in
Ref.\,\cite{DelDuca:2001eu:2001fn}, where the full top-mass dependence
was taken into account. The \nlo{} terms were evaluated in the heavy-top
limit~\cite{Campbell:2006xx}.

The interference of this gluon fusion contribution to $H+2$jets with the
\wbf{} signal at \nlo{} also leads to a net color exchange among the
scattering partons and could diminish the virtues of the actual \wbf{}
signal. However, due to peculiar cancellations, the overall effect of
these terms is negligibly small~\cite{Andersen:2006ag,Andersen:2007mp}.

The subject of this paper is the investigation and classification of
color-exchange terms arising as radiative corrections to the fusion of
weak gauge bosons. We identify a gauge-invariant, finite set of diagrams
that shall allow us to provide an estimate of the size of such terms and
thus their influence on the extraction of the \wbf{} signal from
experimental data.

The remainder of the paper is structured as follows: in
\sct{sec::classification}, we list the various \nnlo{} contributions to
the \wbf{} process, identifying the class of diagrams that is relevant
for our discussion, \sct{sec::calc} describes technical aspects of the
calculation, and \sct{sec::results} contains our results. The
conclusions are presented in \sct{sec::conclusions}.

\section{Classification of the diagrams}\label{sec::classification}

\begin{figure}
  \begin{center}
    \begin{tabular}{ccc}
      \includegraphics[width=.2\textwidth]{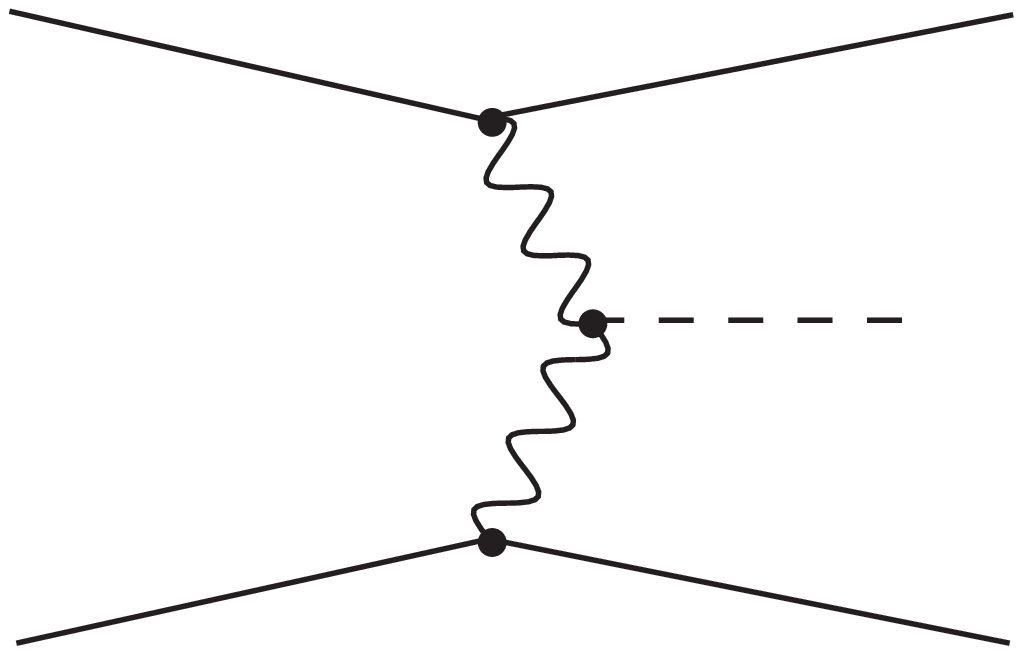} &\qquad\qquad
      \includegraphics[width=.2\textwidth]{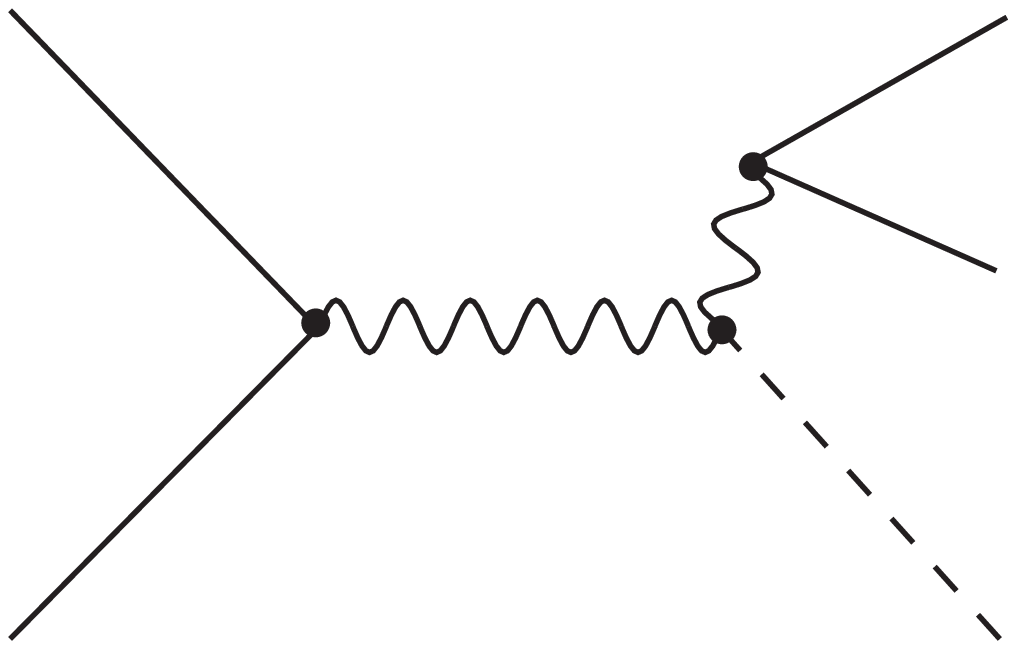}
       &\qquad\qquad
      \includegraphics[width=.2\textwidth]{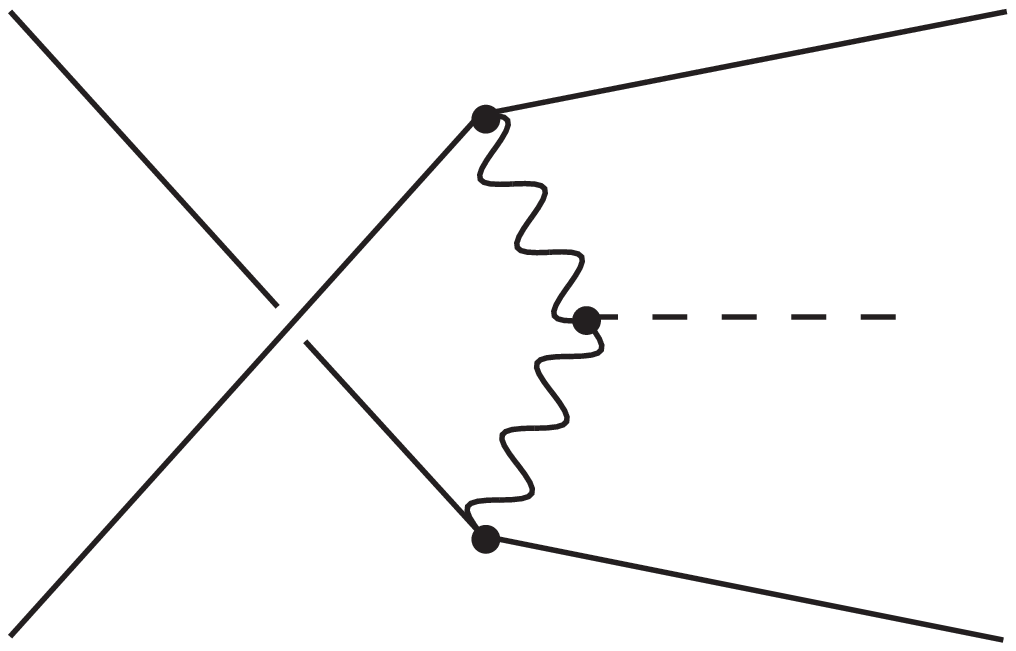}\\
      (a) &\qquad\qquad (b) &\qquad\qquad (c)
    \end{tabular}
    \parbox{.9\textwidth}{
      \caption[]{\label{fig::lo}\sloppy Leading order contributions to
	(a) weak boson fusion and (b) Higgs-Strahlung; (c)~is
	the crossed \lo{} amplitude for \wbf{}. The straight solid lines
	denote quarks ($u,d,c,s,b$), the wiggle lines denote $W$ or $Z$
	bosons, and the dashed line means a Higgs boson.}}
  \end{center}
\end{figure}

Typically, \wbf{} denotes the electro-weak contribution to the process
$pp\to Hjj$ that does not involve resonant weak gauge bosons. A
tree-level example is shown in \fig{fig::lo}\,(a). The contribution
involving resonant weak gauge bosons, \fig{fig::lo}\,(b), is usually
referred to as Higgs-Strahlung. At tree-level, this distinction is
theoretically well-defined. Experimentally, the two contributions can be
distinguished by appropriate cuts: for example, in the \wbf{} process,
the two outgoing jets are typically produced with a much larger
separation in rapidity than in Higgs-Strahlung where they arise from the
decay of a massive gauge boson. At one-loop, on the other hand, the
distinction between \wbf{} and Higgs-Strahlung is less obvious, in
particular for the purely electro-weak corrections. Therefore, in order
to obtain the effects on the \wbf{} process,
Ref.\,\cite{Ciccolini:2007jr:2007ec} calculated the $\order{\alpha}$ and
$\order{\alpha_s}$ corrections to the process $pp\to Hjj$ using the full
set of diagrams and applied the cuts defined for the isolation of the
\wbf{} contribution.

\begin{figure}
  \begin{center}
    \begin{tabular}{cc}
      \includegraphics[width=.2\textwidth]{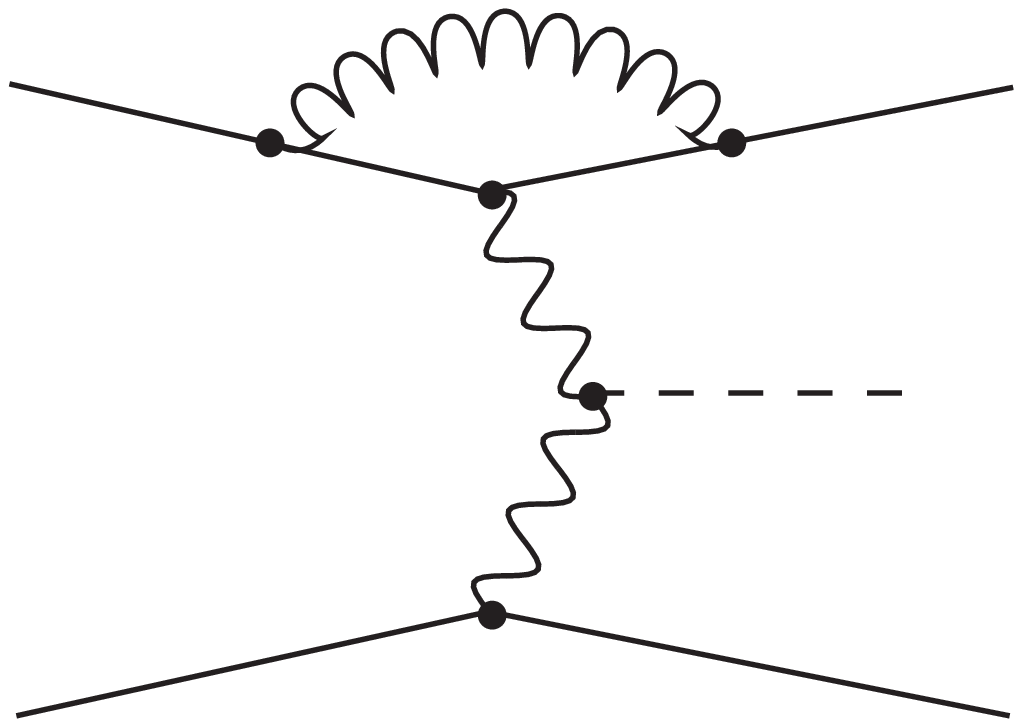} &\qquad
      \includegraphics[width=.2\textwidth]{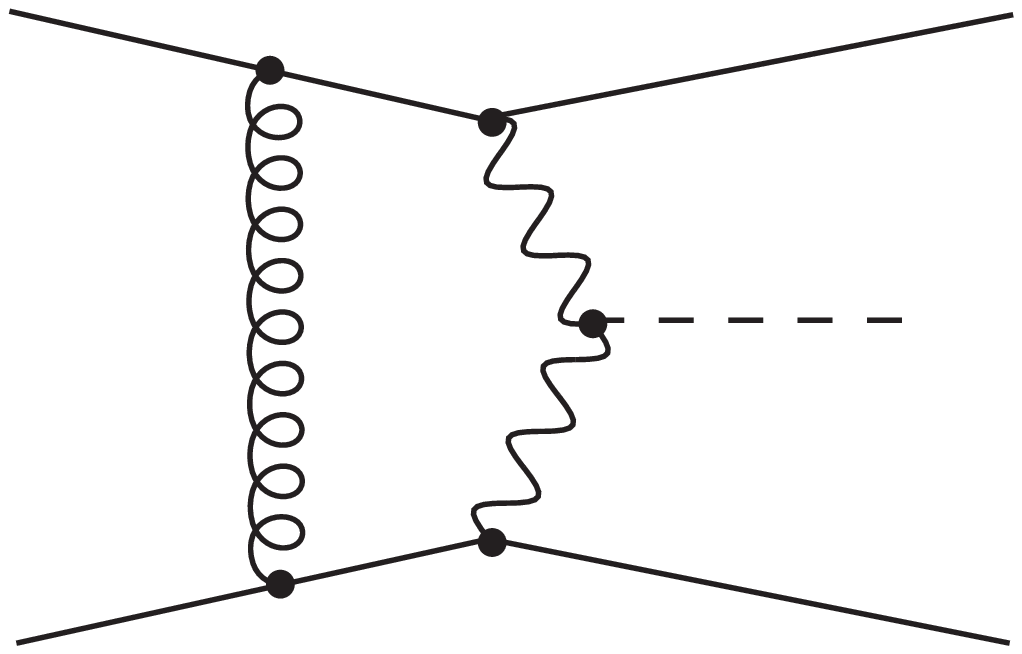}\\
      (a) &\qquad (b)
    \end{tabular}
    \parbox{.9\textwidth}{
      \caption[]{\label{fig::nlo}\sloppy \nlo{} contributions to \wbf{}:
	(a) \dis{}-like and (b) color-exchange diagrams. Notation for
	the straight and wiggle lines is like in \fig{fig::lo}; spiral
	lines denote gluons.  }}
  \end{center}
\end{figure}

When focussing only on the \qcd{} corrections, a distinction between
\wbf{} and Higgs-Strahlung at \nlo{} is still possible, since these
terms are obtained by dressing the leading order diagrams with virtual
or real gluons. The by far dominant \qcd{} corrections to the \wbf{}
contribution come from \dis{}-like
terms~\cite{Han:1992hr,Figy:2003nv,Berger:2004pc,Ciccolini:2007jr:2007ec},
where the gluon
modifies the $q\bar qV$ vertex, see \fig{fig::nlo}\,(a). Diagrams where
the gluon connects the two quark lines, \fig{fig::nlo}\,(b), give no
contribution when interfered with the uncrossed \lo{} amplitude because
of a vanishing color factor.  On the other hand, the interference with
the {\it crossed} leading order amplitude, \fig{fig::lo}\,(c), is
extremely small due to the strong forward-tendency of the outgoing
jets~\cite{Ciccolini:2007jr:2007ec}.

\begin{figure}
  \begin{center}
    \begin{tabular}{cccc}
      &
      \includegraphics[width=.2\textwidth]{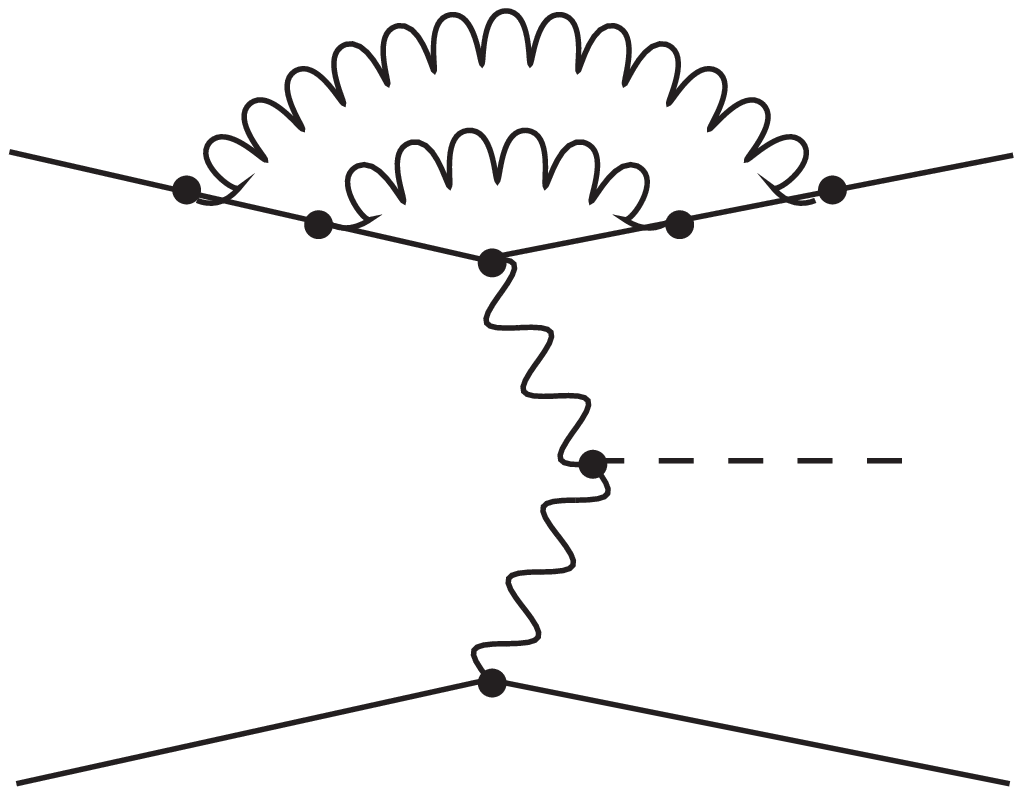} &\qquad
      \raisebox{-.5em}{%
	\includegraphics[width=.2\textwidth]{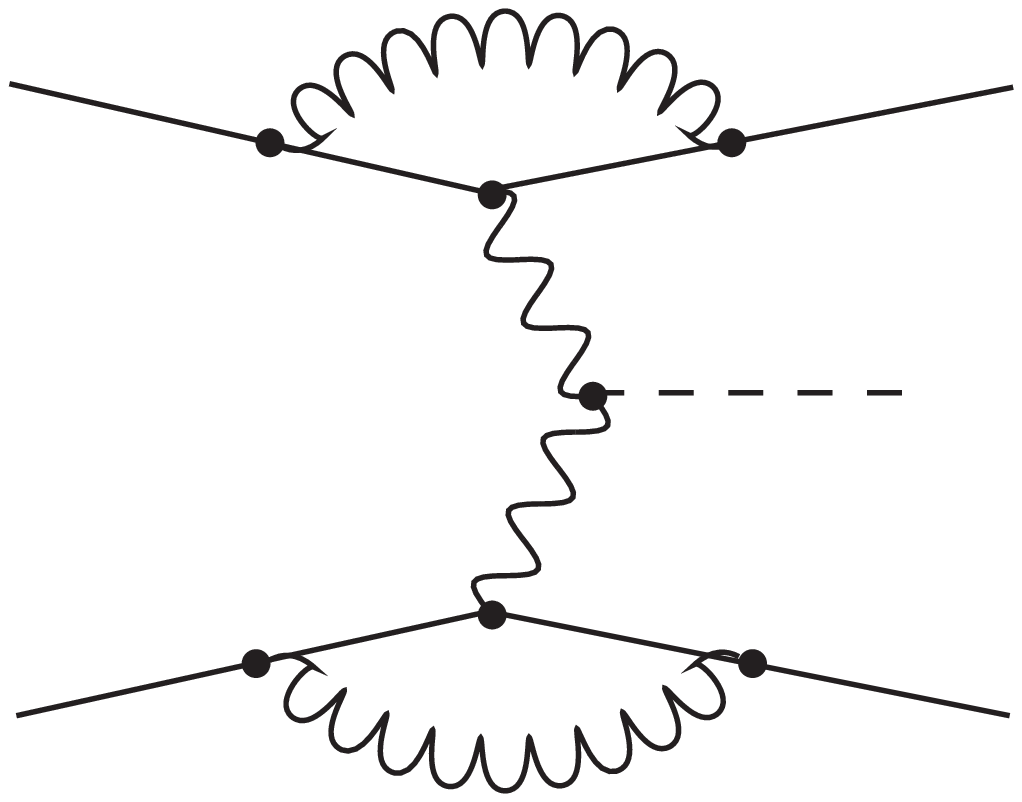}} & \\[.5em]
      & (a) &\qquad (b) &\\[1em]
      \includegraphics[width=.2\textwidth]{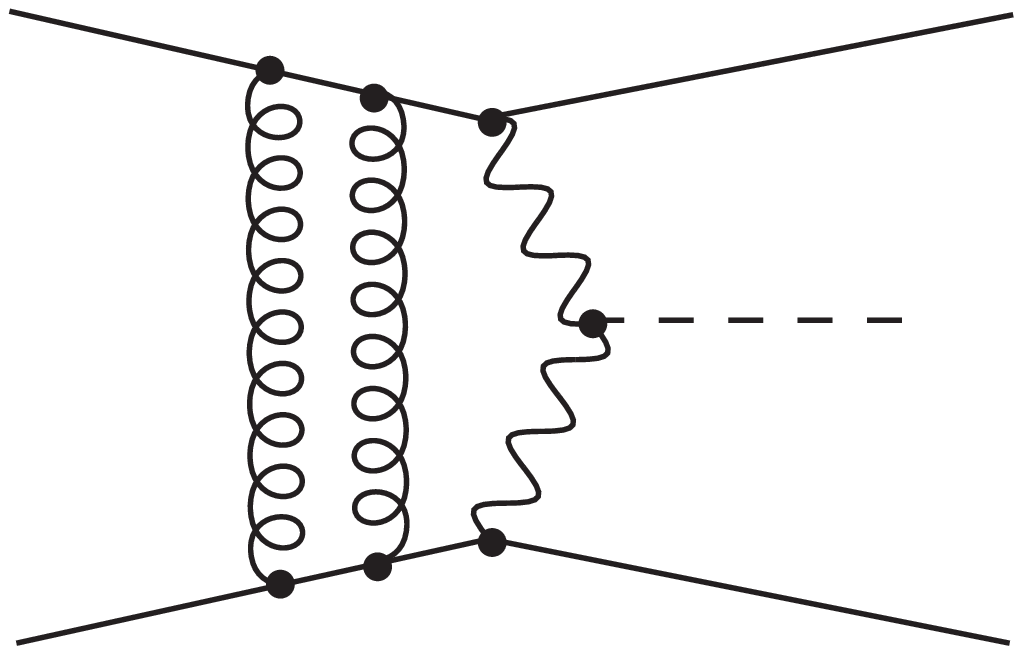} &
      \raisebox{2.5em}{$\star$}
      \raisebox{5.4em}{\includegraphics[width=.2\textwidth,angle=180]{%
	  figs/wbf0.eps}}
      &\qquad
      \includegraphics[width=.2\textwidth]{figs/wbf1nlo.eps} &
      \raisebox{2.5em}{$\star$}
      \raisebox{5.4em}{\includegraphics[width=.2\textwidth,angle=180]{%
	  figs/wbf1nlo.eps}} \\[.5em]
      \multicolumn{2}{c}{(c)} &
      \multicolumn{2}{c}{\qquad(d)}
    \end{tabular}
    \parbox{.9\textwidth}{
      \caption[]{\label{fig::nnlo}\sloppy \nnlo{} \qcd{} contributions
	to \wbf{}: (a), (b) \dis{}-like diagrams and (c), (d)
	color-exchange contributions. Notation is like in
	\fig{fig::nlo}.  }}
  \end{center}
\end{figure}

\begin{figure}
  \begin{center}
    \begin{tabular}{cc}
      \includegraphics[width=.2\textwidth]{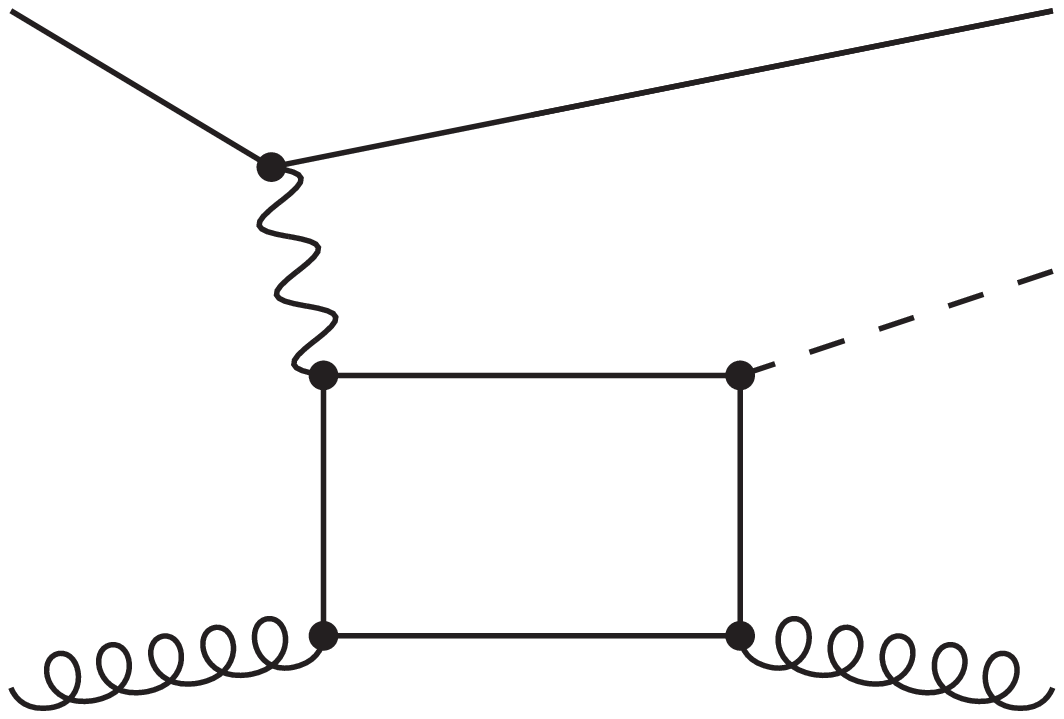} &\qquad
      \includegraphics[width=.2\textwidth]{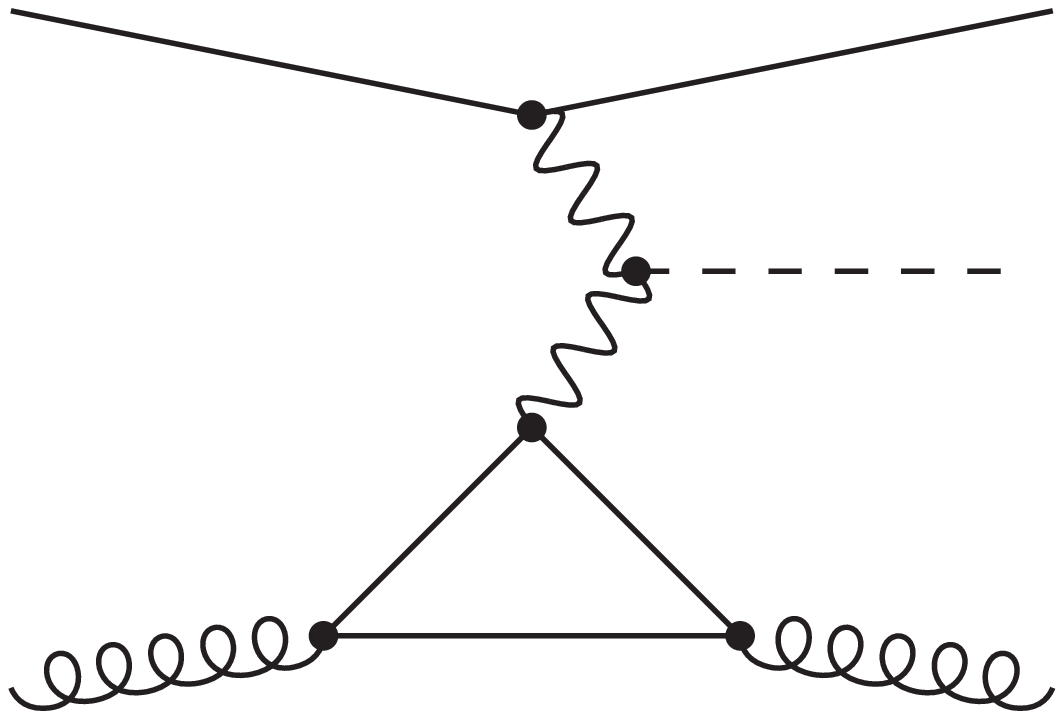}\\[.5em]
      (a) &\qquad (b)
    \end{tabular}
    \parbox{.9\textwidth}{
      \caption[]{\label{fig::quarkloop}\sloppy Diagrams at
	$\order{\alpha^3\alpha_s^2}$ which involve closed quark loops.
	The notation is as in \fig{fig::nlo}, except that the wiggle
	lines can only be $Z$ bosons here, and the particle in the loop can
	be a top quark.}}
  \end{center}
\end{figure}

\begin{figure}
  \begin{center}
    \begin{tabular}{c}
      \begin{tabular}{ccc}
	\includegraphics[width=.2\textwidth]{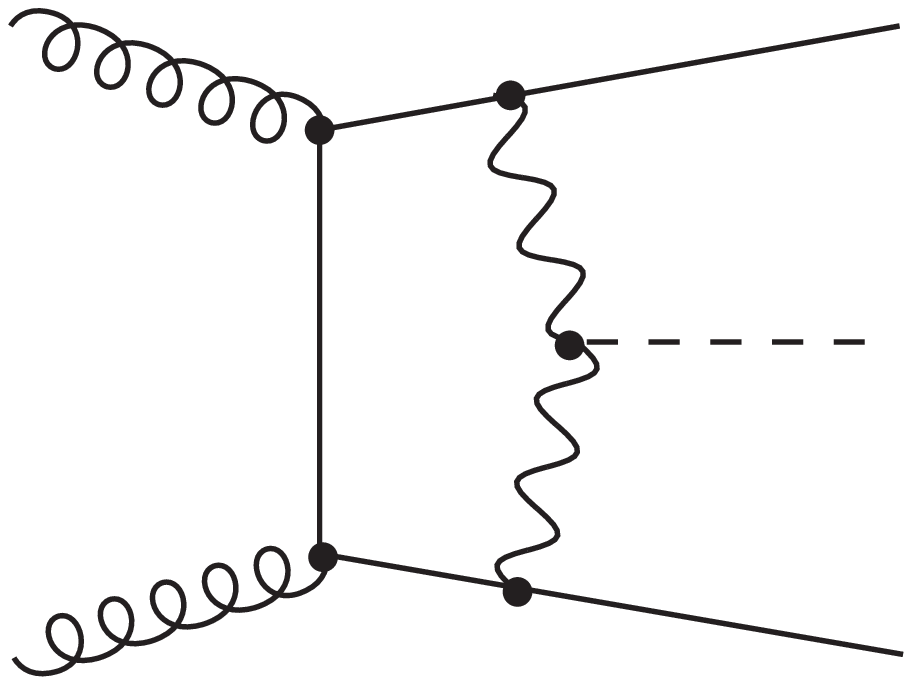} &\qquad
	\includegraphics[width=.2\textwidth]{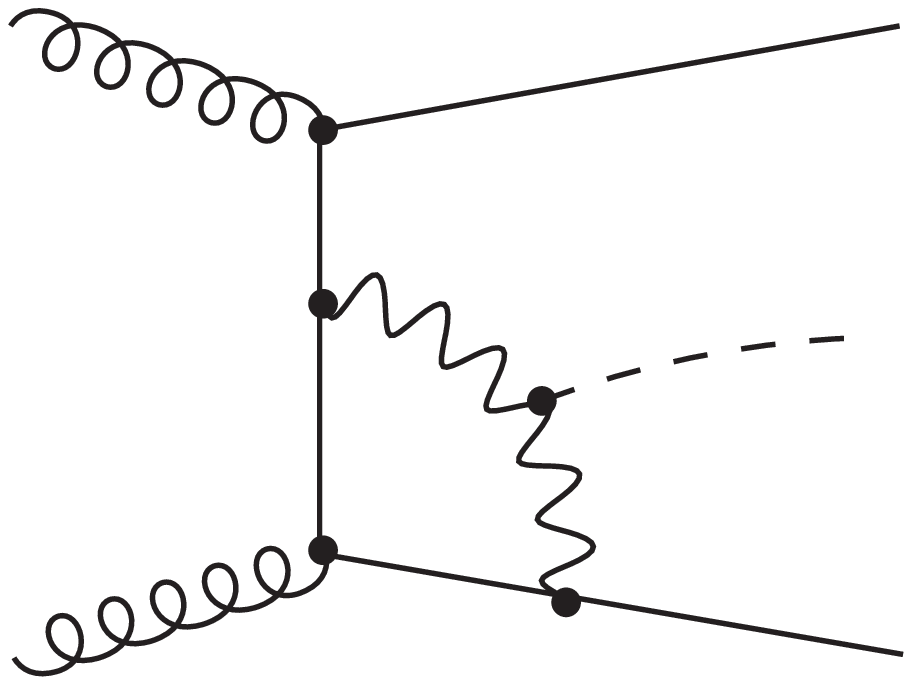} &\qquad
	\includegraphics[width=.2\textwidth]{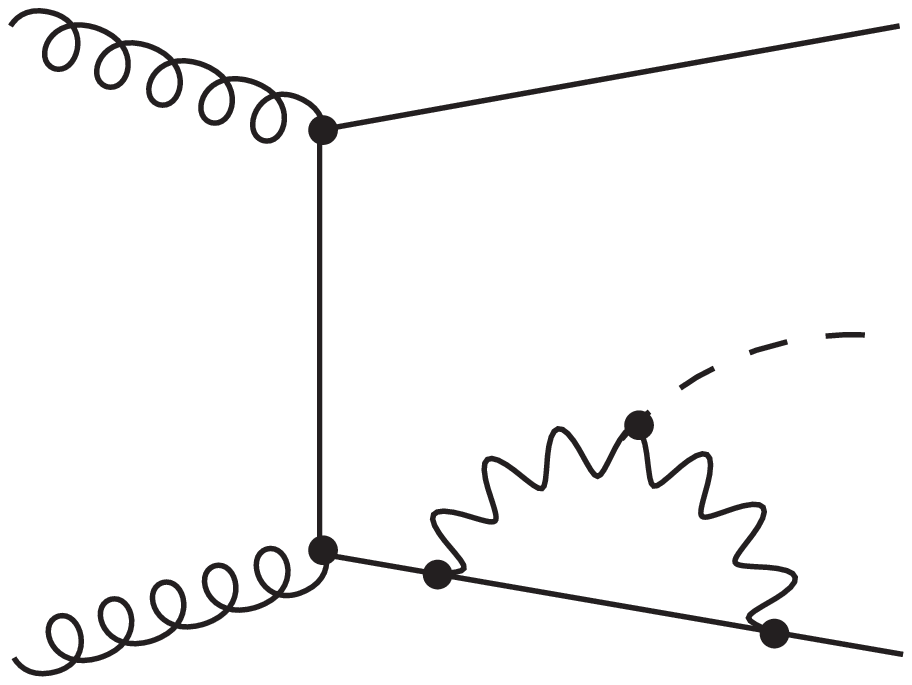} \\
	(a)\qquad & (b)\qquad & (c)\\[1em]
      \end{tabular}
      \\
      \begin{tabular}{ccc}
	\includegraphics[width=.2\textwidth]{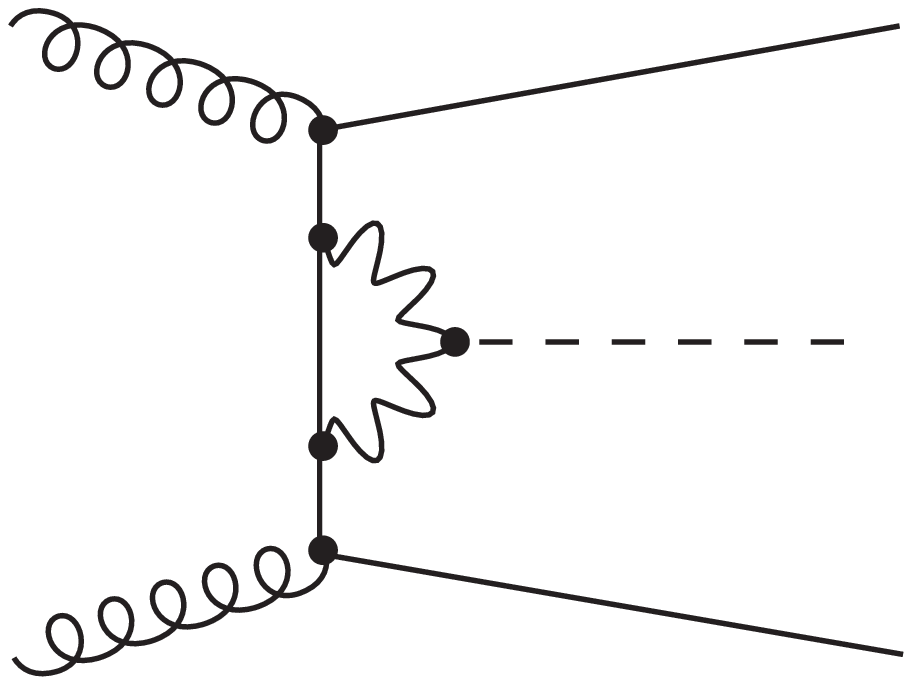} &\qquad
	\raisebox{.5em}{\includegraphics[width=.2\textwidth]{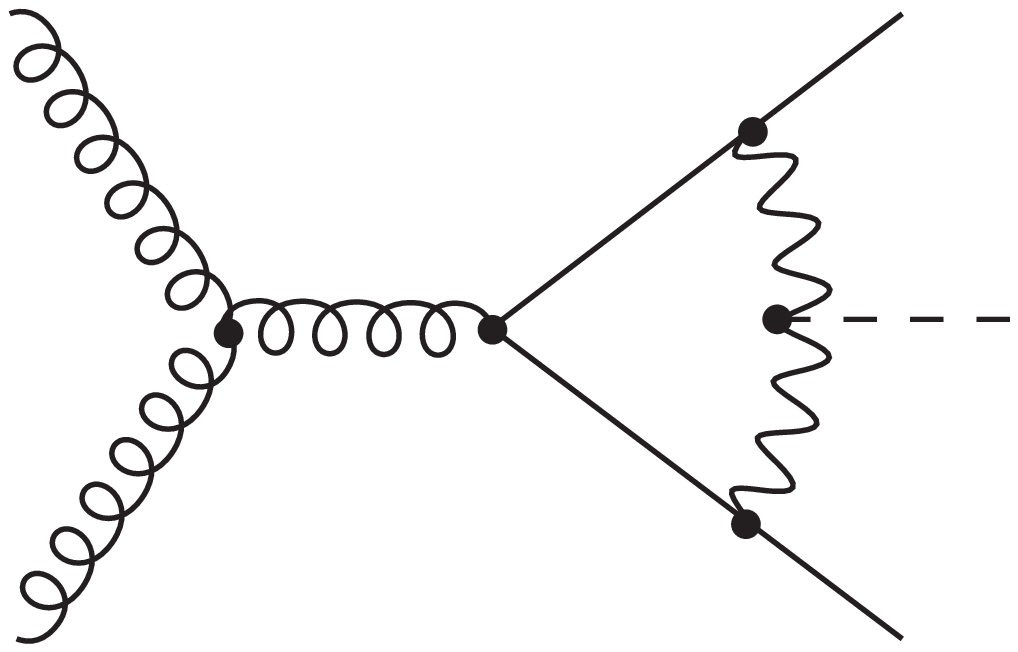}}
	&\qquad
	\raisebox{.5em}{\includegraphics[width=.2\textwidth]{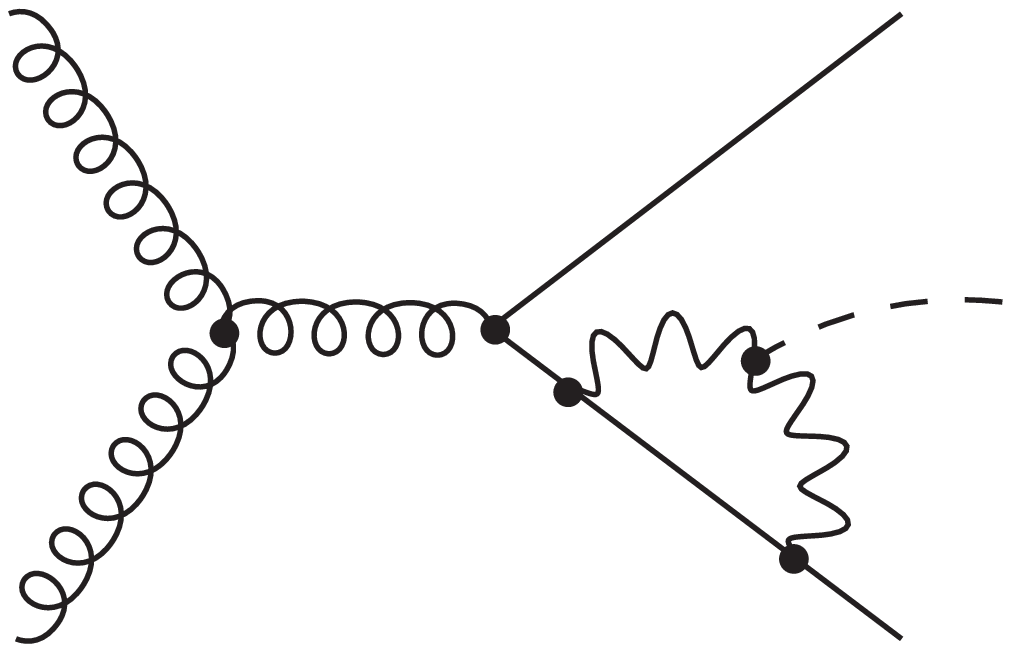}} \\
      (d) &\qquad (e) &\qquad (f)
      \end{tabular}
    \end{tabular}
    \parbox{.9\textwidth}{
      \caption[]{\label{fig::ggwbf}\sloppy
	\nnlo{} contributions to \wbf{} that contain only a single quark
	line. Shown are only the diagrams with purely gluonic initial
	state. $qg$, $\bar qg$, and $q\bar q$ initial state
	contributions are obtained from them by crossing the external
	quark and gluon legs. Notation is like in \fig{fig::nlo}.
        }}
  \end{center}
\end{figure}

This paper addresses the \nnlo{} \qcd{} corrections to the \wbf{}
process, i.e., terms of order $\alpha^3\alpha_s^2$ to $pp\to Hjj$ which do
{\it not} involve resonant weak gauge bosons.  Among them, we may still
distinguish various contributions:
\begin{itemize}
\item \dis{}-like terms, i.e.\ those that merely involve corrections to
the $q\bar qV$ vertex, see \fig{fig::nnlo}\,(a),(b). Due to this
similarity and because of the structure of the \nlo{} results, we expect
them to be similar in size as the \nnlo{} corrections to the \dis{}
process themselves~\cite{Zijlstra:1992qd}.
\item Diagrams involving gluon exchange between the two quark lines. Due
to the possibility of two gluons forming a color-singlet state, the
interference with the {\it uncrossed} amplitude is non-zero, see
\fig{fig::nnlo}\,(c),(d).  Technically, the double gluon exchange
diagram (c) is probably the most difficult one when aiming for the full
\nnlo{} result. Note, however, that the net color exchange is still
equal to zero in this case. Thus, we do not expect this contribution to
significantly increase the hadronic activity at central rapidities. It
goes without saying that in order to arrive at infra-red finite results,
all the above contributions require the calculation of single and double
real gluon radiation.
\item Diagrams involving closed quark loops, \fig{fig::quarkloop}.  Due
  to the smallness of the Yukawa couplings, only the top quark occurs in
  the loop of diagram (a). In diagram (b), on the other hand, only the
  third generation quarks survive the Furry theorem. Neither (a) nor (b)
  involves color exchange among the scattering partons.  Note that
  crossing the external partons may lead to diagrams with a resonant
  weak boson and thus are not counted as \wbf{}. In particular, the
  diagrams where both initial state particles are gluons were calculated
  in Ref.\,\cite{Brein:2003wg}.
\item Diagrams that involve only a single quark line, to be referred to
  as \sql{} diagrams in what follows. A generic set is shown in
  \fig{fig::ggwbf}. The other terms of this class are obtained by simply
  crossing the external quarks and gluons, thus leading also to $qg$,
  $\bar qg$, and $\bar qq$ initial states.
\end{itemize}

Thus, the only diagrams involving a net color exchange among the
scattering partons are \fig{fig::nnlo}\,(d) and obvious variants, plus
the \sql{} contribution. In order to get a handle on the size of the
color exchange, in this paper we shall focus on the latter, i.e., the
diagrams of \fig{fig::ggwbf}. They form a gauge-invariant and {\abbrev
UV}-finite set and are {\abbrev IR} finite without taking any real
radiation into account. What also motivates the calculation of these
diagrams is that they are the only ones through \nnlo{} that involve
purely gluonic initial states.  Recalling that the gluon luminosity at
the \lhc{} is very large, one may expect their numerical impact to be
rather sizable.

\section{Calculation}\label{sec::calc}

The Feynman diagrams were generated using {\tt
  FeynArts}~\cite{Kublbeck:1990xc,Hahn:2000kx}, taking into account the
five light quark flavors, both in the initial and in the final state.
In principle, the quark line in \fig{fig::ggwbf} could also be the top
quark (for the $gg$ channel), but we expect this contribution to be
very small due to the large $x$ values required for the incoming
gluons and the reduced available phase space.

The Feynman diagrams are evaluated using standard techniques.  The
amplitudes are simplified with {\tt FormCalc}~\cite{Hahn:1998yk} and the
results in terms of Weyl-spinor chains and coefficients containing the
tensor one-loop integrals have been translated to {\tt C++} code for the
numerical evaluation.  The one-loop integrals are reduced to a set of
standard integrals numerically.  The 5-point integrals are written in
terms of 4-point functions following Ref.~\cite{Denner:2002ii}, where a
method for a direct reduction is described that avoids leading inverse
Gram determinants and the associated numerical instabilities.  The
remaining tensor integrals are recursively reduced to scalar integrals
with the Passarino--Veltman algorithm \cite{Passarino:1978jh} for
non-exceptional phase-space points.  In the exceptional phase-space
regions the reduction of the 3- and 4-point tensor integrals is
performed using the methods of Ref.~\cite{Denner:2005nn} which allow for
a numerically stable evaluation. The scalar integrals are calculated
using the results of Refs.~\cite{'t Hooft:1978xw, Denner:1991qq}.  For
the numerical evaluation of the one-loop integrals we use a library by
A.~Denner implementing the methods of Refs.~\cite{Denner:2002ii,
Denner:2005nn}.

The phase-space integration is performed with Monte Carlo techniques
using the adaptive multi-dimensional integration program {\tt Vegas}
\cite{Lepage:1977sw}.

The correctness of our results was checked on the one hand by confirming
their gauge invariance in the usual way, i.e.\ by replacing the
external gluon polarization vector by the incoming momentum. On the
other hand, the major part of the diagrams was independently calculated
completely within {\tt FeynArts}/{\tt FormCalc}/{\tt
LoopTools}~\cite{Kublbeck:1990xc,Hahn:2000kx,Hahn:1998yk}. Of course,
complete agreement was found among the two calculations.

Note that some of the diagrams, e.g.~\fig{fig::ggwbf}\,(b), are singular
as one or both of the outgoing partons become soft or collinear to the
incoming parton(s).  This cannot occur if we ask for the typical \wbf{}
signal which involves two hard jets, however. In fact, our minimal set of cuts
imposed on the outgoing jets is
\begin{equation}
\begin{split}
p_{Tj} > 20\,\mbox{GeV}\,,\qquad
|\eta_j|<5\,,\qquad
R>0.6\,,\qquad j\in\{1,2\}\,,
\label{eq::mincuts}
\end{split}
\end{equation}
where $p_{Tj}$ and $\eta_j$ are the transverse momentum and the
pseudo-rapidities of the final state jets, respectively.
$R$ is the separation of the jets in the $\eta$--$\phi$ plane,
\begin{equation}
\begin{split}
R= \sqrt{(\Delta \eta)^2 + (\Delta \phi)^2}\,,\qquad \Delta\eta = \eta_1
- \eta_2\,,\qquad \Delta\phi = \phi_1 - \phi_2\,,
\end{split}
\end{equation}
where $\phi_j$
is the azimuthal angle of the jet.
\eqn{eq::mincuts} ensures that the events contain two well-separated
hard jets at not too large rapidities.

The virtue of the \wbf{} process is that a set of additional cuts on the
outgoing jets allows for a significant improvement of the
signal-to-background ratio, where ``background'' also includes Higgs
production by other mechanisms than \wbf{}, in particular by processes
involving resonant gauge bosons. These so-called \wbf{} cuts are given
by \eqn{eq::mincuts} plus
\begin{equation}
\begin{split}
\eta_1\cdot \eta_2<0\,,\qquad
|\Delta\eta|>4.2\,,\qquad
m_{jj}> 600\,\mbox{GeV}\,,
\label{eq::wbfcuts}
\end{split}
\end{equation}
where $m_{jj}$ is the invariant mass of the two-jet system.
These cuts allow only for events in which the jets lie in opposite
hemispheres of the detector, are separated by a significant rapidity
gap, and have a large invariant mass.

\section{Results}\label{sec::results}

Since the \sql{} diagrams are a \nnlo{} contribution to \wbf{} we use
the {\abbrev NNLO MRST2004} parton distributions \cite{Martin:2004ir} and a
2-loop running $\alpha_s$. As the scale in the evaluation of both the
{\abbrev PDF}s and $\alpha_s$ we use $\mu = m_H$. The input parameters for the
electro-weak sector are $\alpha = 1/137.036$, $m_W = 80.423 \GeV$, $m_Z
= 91.1876 \GeV$ and $\sin^2\theta_W = 0.222$.

\begin{figure}
  \begin{center}
    \begin{tabular}{c}
      \includegraphics[bb=90 250 480 560,%
	width=.55\textwidth]{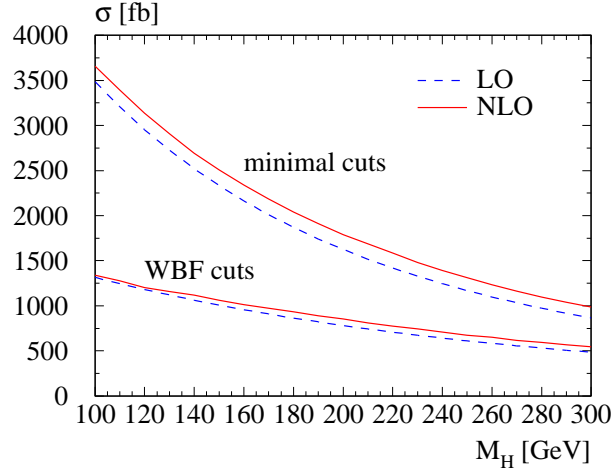}
    \end{tabular}
    \parbox{.9\textwidth}{
      \caption[]{\label{fig::lonlo}\sloppy (a) Total \wbf{} cross
	section at the \lhc{} at \lo{} and \nlo{}, both with the minimal
	set of cuts, \eqn{eq::mincuts}, and with \wbf{}
	cuts. Curves obtained using Ref.\,\cite{vbfnlo-1.2}.}}
  \end{center}
\end{figure}

For comparison, in \fig{fig::lonlo} we show the total cross section for
\wbf{} at the \lhc{} at \lo{} and \nlo{} \qcd{}, both with the minimal
set of cuts, \eqn{eq::mincuts}, and with \wbf{} cuts,
\eqn{eq::wbfcuts}. The curves were produced using the program {\tt
vbfnlo}~\cite{vbfnlo-1.2}.  One observes that the \nlo{} corrections are
of the order of 10\%, and that the \wbf{} cuts reduce the cross section by
roughly a factor 2-3.

\begin{figure}
  \begin{center}
    \begin{tabular}{cc}
      \includegraphics[bb=40 400 260 640,%
	width=.45\textwidth]{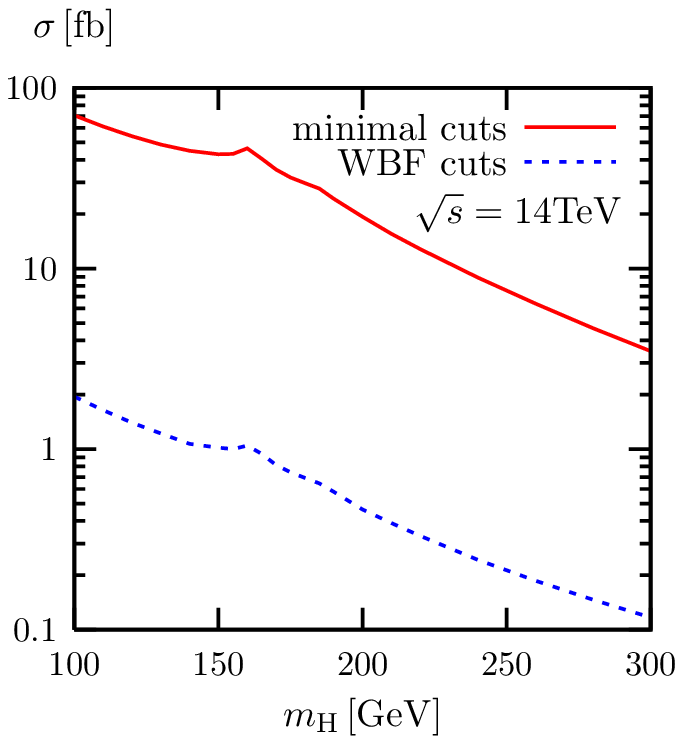} &
      \includegraphics[bb=40 400 260 640,%
	width=.45\textwidth]{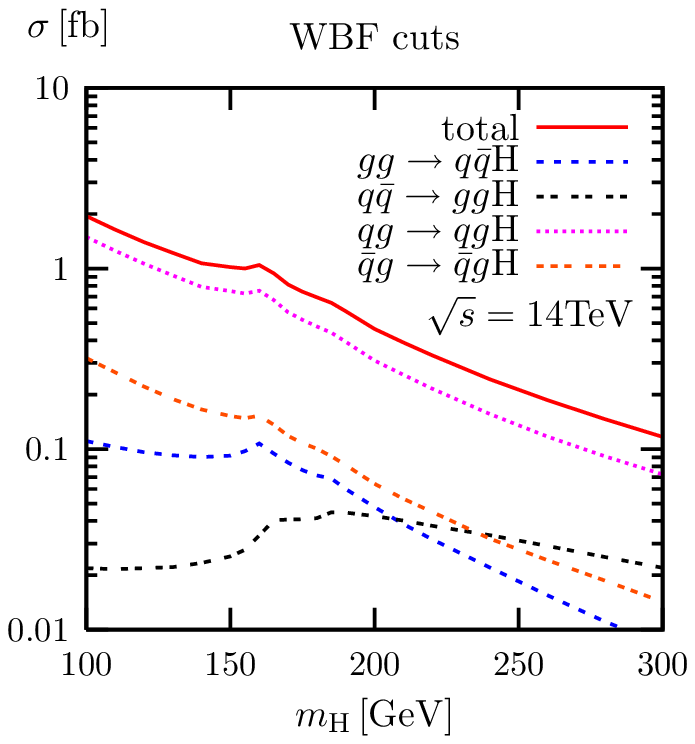}\\
      (a) & (b)
    \end{tabular}
    \parbox{.9\textwidth}{
      \caption[]{\label{fig::sigma}\sloppy (a) Total cross section at
	the \lhc{} for the \sql{} contribution of \fig{fig::ggwbf} as a
	function of the Higgs mass $m_H$. Solid: with the minimal set of
	cuts, \eqn{eq::mincuts}; dashed: with \wbf{} cuts. (b)
	Individual contributions to the total cross section at the
	\lhc{} when \wbf{} cuts are applied.}}
  \end{center}
\end{figure}

\fig{fig::sigma} shows the cross section due to the \sql{} contribution
of \fig{fig::ggwbf}. When minimal cuts are applied, the overall
magnitude amounts to roughly 2\% of the \lo{} terms for
$M_H=100\GeV$. However, the fall-off of the \sql{} contribution with
increasing $M_H$ is much steeper than for the conventional \wbf{}
process (note the logarithmic scale).  \fig{fig::sigma}\,(b) shows the
individual contributions from $gg$, $qg$, $\bar qg$, and $q\bar q$
initial states as they are obtained by crossing the external partons in
\fig{fig::ggwbf}. It is quite remarkable that the $qg$ component
dominates all the other channels by almost an order of magnitude.

\begin{figure}
  \begin{center}
    \begin{tabular}{cc}
      \includegraphics[bb=40 400 260
      640,width=.45\textwidth]{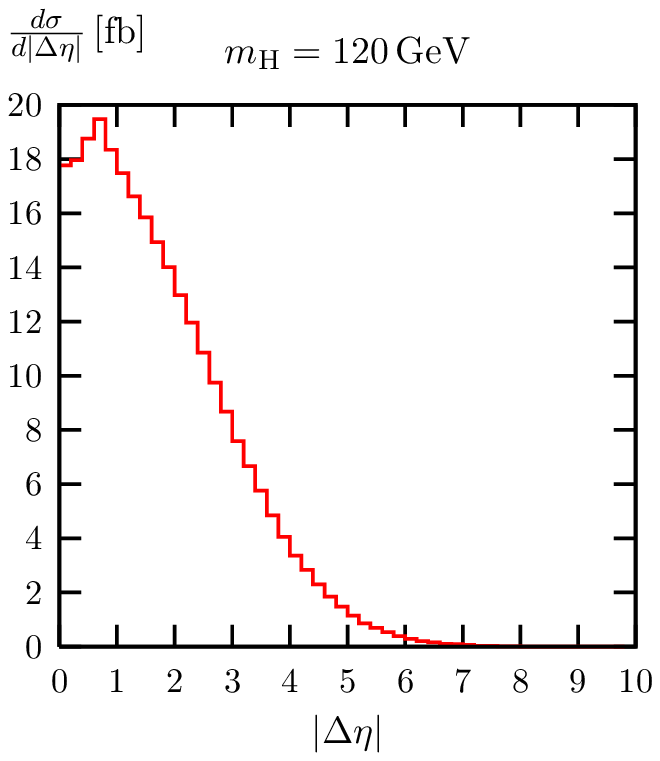} &
      \includegraphics[bb=40 400 260
      640,width=.45\textwidth]{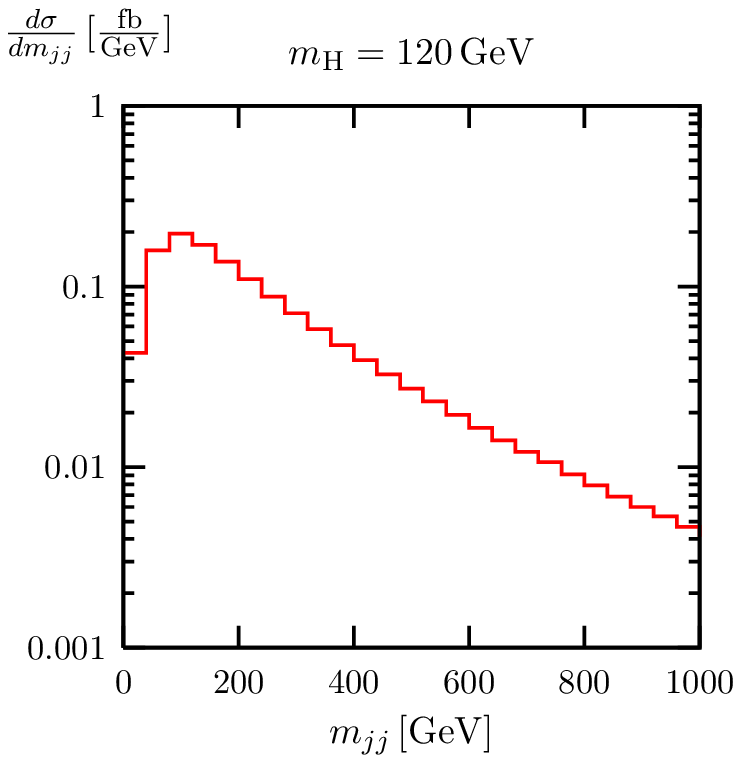}\\
      (a) & (b)
    \end{tabular}
    \parbox{.9\textwidth}{
      \caption[]{\label{fig::distrib}\sloppy
	Kinematical distributions of the final state jets for the \sql{}
	contribution using minimal cuts, \eqn{eq::mincuts}.}}
  \end{center}
\end{figure}

\wbf{} cuts suppress the \sql{} contribution by roughly a factor of 30,
meaning that its effect on the \wbf{} signal is completely negligible.
The reason for this large suppression becomes clear once the relevant
kinematical distributions are considered in more
detail. \fig{fig::distrib}\,(a) and (b) show the distribution for the
separation of the two jets when produced through \sql{} diagrams and the
invariant mass of the two-jet system, respectively. The curves clearly
show that the bulk of the events lies below the cuts of
\eqn{eq::wbfcuts}. In particular the cut on $|\Delta\eta|$ is extremely
effective in removing this potentially dangerous component of the \wbf{}
process, thereby preserving its promising perspective for Higgs studies
at the \lhc{}.

\section{Conclusions}\label{sec::conclusions}

\wbf{} is a remarkable channel in the sense that the Higgs decay
products are found in a region of the detector where hadronic activity
is rather low. We have shown that for a potential source of such
hadronic activity that occurs at \nnlo{}, among them the purely
gluon-initiated contribution, this statement remains true, because it is
efficiently suppressed by the usual \wbf{} cuts. In combination with
recent results on interference effects with the gluon fusion
process~\cite{Andersen:2007mp}, this is a reassuring observation
concerning the usefulness of the \wbf{} process at the \lhc{}.\\[2em]

{\bf Acknowledgments.}\qquad We would like to thank A.~Djouadi for
initially pointing out this problem to us. We also thank him and
T.~Binoth for useful comments on the manuscript.  R.H.\ acknowledges the
hospitality of the {\it Galileo Galilei Institute} in Florence where
part of this work was carried out.  This work was financially supported
by {\it Deutsche Forschungsgemeinschaft} under grants HA~2990/2-1 and
HA~2990/3-1. The work of M.W. was supported in part by the National
Science Foundation under grant NSF-PHY-0547564.

\def\app#1#2#3{{\it Act.~Phys.~Pol.~}\jref{\bf B #1}{#2}{#3}}
\def\apa#1#2#3{{\it Act.~Phys.~Austr.~}\jref{\bf#1}{#2}{#3}}
\def\annphys#1#2#3{{\it Ann.~Phys.~}\jref{\bf #1}{#2}{#3}}
\def\cmp#1#2#3{{\it Comm.~Math.~Phys.~}\jref{\bf #1}{#2}{#3}}
\def\cpc#1#2#3{{\it Comp.~Phys.~Commun.~}\jref{\bf #1}{#2}{#3}}
\def\epjc#1#2#3{{\it Eur.\ Phys.\ J.\ }\jref{\bf C #1}{#2}{#3}}
\def\fortp#1#2#3{{\it Fortschr.~Phys.~}\jref{\bf#1}{#2}{#3}}
\def\ijmpc#1#2#3{{\it Int.~J.~Mod.~Phys.~}\jref{\bf C #1}{#2}{#3}}
\def\ijmpa#1#2#3{{\it Int.~J.~Mod.~Phys.~}\jref{\bf A #1}{#2}{#3}}
\def\jcp#1#2#3{{\it J.~Comp.~Phys.~}\jref{\bf #1}{#2}{#3}}
\def\jetp#1#2#3{{\it JETP~Lett.~}\jref{\bf #1}{#2}{#3}}
\def\jhep#1#2#3{{\small\it JHEP~}\jref{\bf #1}{#2}{#3}}
\def\mpl#1#2#3{{\it Mod.~Phys.~Lett.~}\jref{\bf A #1}{#2}{#3}}
\def\nima#1#2#3{{\it Nucl.~Inst.~Meth.~}\jref{\bf A #1}{#2}{#3}}
\def\npb#1#2#3{{\it Nucl.~Phys.~}\jref{\bf B #1}{#2}{#3}}
\def\nca#1#2#3{{\it Nuovo~Cim.~}\jref{\bf #1A}{#2}{#3}}
\def\plb#1#2#3{{\it Phys.~Lett.~}\jref{\bf B #1}{#2}{#3}}
\def\prc#1#2#3{{\it Phys.~Reports }\jref{\bf #1}{#2}{#3}}
\def\prd#1#2#3{{\it Phys.~Rev.~}\jref{\bf D #1}{#2}{#3}}
\def\pR#1#2#3{{\it Phys.~Rev.~}\jref{\bf #1}{#2}{#3}}
\def\prl#1#2#3{{\it Phys.~Rev.~Lett.~}\jref{\bf #1}{#2}{#3}}
\def\pr#1#2#3{{\it Phys.~Reports }\jref{\bf #1}{#2}{#3}}
\def\ptp#1#2#3{{\it Prog.~Theor.~Phys.~}\jref{\bf #1}{#2}{#3}}
\def\ppnp#1#2#3{{\it Prog.~Part.~Nucl.~Phys.~}\jref{\bf #1}{#2}{#3}}
\def\rmp#1#2#3{{\it Rev.~Mod.~Phys.~}\jref{\bf #1}{#2}{#3}}
\def\sovnp#1#2#3{{\it Sov.~J.~Nucl.~Phys.~}\jref{\bf #1}{#2}{#3}}
\def\sovus#1#2#3{{\it Sov.~Phys.~Usp.~}\jref{\bf #1}{#2}{#3}}
\def\tmf#1#2#3{{\it Teor.~Mat.~Fiz.~}\jref{\bf #1}{#2}{#3}}
\def\tmp#1#2#3{{\it Theor.~Math.~Phys.~}\jref{\bf #1}{#2}{#3}}
\def\yadfiz#1#2#3{{\it Yad.~Fiz.~}\jref{\bf #1}{#2}{#3}}
\def\zpc#1#2#3{{\it Z.~Phys.~}\jref{\bf C #1}{#2}{#3}}
\def\ibid#1#2#3{{ibid.~}\jref{\bf #1}{#2}{#3}}
\def\otherjournal#1#2#3#4{{\it #1}\jref{\bf #2}{#3}{#4}}

\newcommand{\jref}[3]{{\bf #1} (#2) #3}
\newcommand{\bibentry}[4]{#1, #3\ifthenelse{\equal{#4}{}}{}{, }#4.}
\newcommand{\hepph}[1]{{\tt hep-ph/#1}}
\newcommand{\arxiv}[1]{{\tt arXiv:#1}}

\end{document}